\documentclass[preprint,showpacs,amsmath,amssymb]{revtex4}
\usepackage{graphicx}% Include figure files
\usepackage{dcolumn}% Align table columns on decimal point
\usepackage{bm}% bold math

\newcommand{\GMA}{$\rm{Ga_{1-x}Mn_{x}As}$}
\newcommand{\RA}{$R_A$}
\newcommand{\RP}{$R_P$}
\newcommand{\PG}{$P_G$}
\newcommand{\PM}{$P_M$}
\newcommand{\da}{$d_{\rm{AlAs}}$}
\newcommand{\greeksym}[1]{{\usefont{U}{psy}{m}{n}#1}}
\newcommand{\umu}{\mbox{\greeksym{m}}}

\begin{document}
\title{Spin-polarized Tunneling in Hybrid Metal-Semiconductor Magnetic Tunnel Junctions}
\author{S. H.\ Chun}\email{schun@psu.edu}
\author{S. J. Potashnik}
\author{K. C. Ku}
\author{P. Schiffer}
\author{N.\ Samarth}\email{nsamarth@psu.edu}
\affiliation{Department of Physics and Materials Research Institute, The Pennsylvania State University,
University Park PA 16802}

\begin{abstract}
We demonstrate efficient spin-polarized tunneling between a ferromagnetic metal and a ferromagnetic semiconductor with highly mismatched conductivities. This is indicated by a large tunneling magnetoresistance (up to 30$\%$) at low temperatures in epitaxial magnetic tunnel junctions composed of a ferromagnetic metal (MnAs) and a ferromagnetic semiconductor (\GMA) separated by a nonmagnetic semiconductor (AlAs). Analysis of the current-voltage characteristics yields detailed information about the asymmetric tunnel barrier. The low temperature conductance-voltage characteristics show a zero bias anomaly and a $\sqrt{V}$~ dependence of the conductance, indicating a correlation gap in the density of states of \GMA. These experiments suggest that MnAs/AlAs heterostructures offer well characterized tunnel junctions for high efficiency spin injection into GaAs. 
\end{abstract}
\maketitle

Fundamental studies of spin-dependent transport and tunneling in metallic ferromagnetic heterostructures have been of critical importance to the development of metallic ``spintronic'' devices for high density information storage.\cite{prinzspintronics} The emerging interest in a {\it semiconductor}-based ``spintronics'' technology has now sparked substantial interest in studies of similar phenomena in semiconductor heterostructures.\cite{spinbook} An important hurdle in this context is the inefficient injection of spin-polarized currents from metallic ferromagnets into semiconductors due to the large mismatch in conductivities.\cite{schmidtPRB00} This problem can be overcome by using either ferromagnetic semiconductors or highly spin-polarized paramagnetic semiconductors for spin injection.\cite{ohnonature99} An alternative solution uses spin-injection from a ferromagnetic metal via a tunnel barrier, so that the conductivity mismatch problem is essentially circumvented by the large contact resistance.\cite{rashbaPRB00} While recent spin injection experiments involving both electrical \cite{johnsonPRL02} and optical \cite{jonkerAPL02} detection schemes suggest the successful demonstration of this concept, extrinsic effects can complicate the correct interpretation of these experiments.\cite{filipPRB00,crowellPRB02}  Rashba has proposed a more direct scheme for measuring spin injection through a tunnel barrier wherein one measures the tunneling magnetoresistance (TMR) between metallic ferromagnetic tunnel contacts that sandwich a semiconductor.\cite{rashbaPRB00} Here, one tunnel barrier serves as the spin injector and the second as the spin detector, and the physics is completely analogous to that of a traditional magnetic tunnel junction (MTJ).\cite{mooderaPRL95} The fabrication of high quality epitaxial metal/semiconductor/metal heterostructures needed for such a scheme presents a difficult materials challenge. For instance, even in the most successful examples of such epitaxial MTJs (MnAs/AlAs/MnAs), the magnetoresistance effects are small ($\approx 1 \%$).\cite{tanakaAPL02}

Here, we demonstrate efficient spin-polarized tunneling in a new class of ``hybrid'' epitaxial MTJs comprised of a ferromagnetic metal (MnAs) and a ferromagnetic semiconductor (\GMA) separated by a nonmagnetic semiconductor (AlAs). In carefully designed heterostructures, we observe a significant TMR ranging up to 30$\%$ at low temperatures. Even with conservative estimates for the spin polarization in the ferromagnetic layers, this observation indicates highly efficient spin injection through the AlAs barrier, suggesting that MnAs/AlAs tunnel contacts offer an attractive scheme for spin injection into GaAs. Although the current experiment is limited to detecting spin injection at temperatures below the relatively low Curie temperature \GMA~ ($T_C = 70$K), the high Curie temperature of MnAs ($T_C = 320$K) allows for future room temperature experiments in different configurations.  We note the low Curie temperature of the \GMA~ layer provides a built-in control experiment since we can measure the TMR in both the paramagnetic and ferromagnetic states of \GMA. 

It is also important to place the current experiments within the context of traditional studies of MTJs.  As is well-known, MTJs typically involve only metallic ferromagnets,\cite{mooderaPRL95} and the physics is well-described by the ``generic'' Julliere model.\cite{julliere} More recent experiments using ferromagnetic semiconductor MTJs\cite{hayashiJCG99,chibaAPL00} have instead needed detailed band structure modeling to properly explain the tunneling.\cite{tanakaPRL01} Both these all-metal or all-semiconductor MTJs necessarily probe tunneling between materials with similar conductivity. In contrast, the MnAs/AlAs/\GMA~ heterostructures studied here have conductivities differing by roughly four orders of magnitude ($\approx$~1\umu $\Omega$-cm for MnAs and $\approx$10 m$\Omega$-cm for \GMA).\cite{berryPRB01,potashnikAPL01} These hybrid systems hence open up a distinct new class of MTJs and have the potential to yield qualitatively new insights into the physics of magnetic tunnel junctions.\cite{liuJCG01}   

The MTJ samples are fabricated by MBE on  p$^+$-GaAs (001) substrates after the growth of a 40 nm-thick p-GaAs buffer layer.  We have studied a wide variety of sample configurations that involve as components the materials GaAs, (Ga,Mn)As, (Ga,Al)As and MnAs, but focus here on a systematic set of 4 samples wherein \GMA~ (x = 0.03, thickness 120 nm), GaAs(thickness 1 nm), AlAs (thickness \da = 1 nm, 2nm, 5 nm and 10nm), GaAs (thickness 1 nm), and MnAs (thickness 45 nm) are grown sequentially at 250$^0$C. We note that -- as in the case of \GMA/AlAs/\GMA~  MTJs \cite{tanakaPRL01} -- the thin GaAs spacer layers placed between the ferromagnetic layers and the tunnel barrier appear to be crucial to the observation of distinct TMR characteristics. Reflection high energy electron diffraction measurements during the growth confirm the epitaxy of MnAs in the ``type-B'' orientation.\cite{tanakaAPL94}  Standard photolithography and wet-etching techniques are used to define 300 \umu m-diameter mesas for vertical transport measurements. Each mesa is etched down into the p-GaAs region, and the DC current-voltage characteristics of a mesa between the top MnAs layer and the back of the p-GaAs substrate are then measured using a four-probe method.  These measurements are carried out in a continuous flow He cryostat over the range 4.2 K - 300 K with an in-plane magnetic field ranging up to 2 kG provided by an electromagnet; additional transport measurements down to 330 mK are carried out in a He 3 cryostat with a superconducting magnet. Finally, magnetization measurements are carried out on 10 mm$^2$ pieces of the unpatterned wafer using a Quantum Design superconducting quantum interference device (SQUID) magnetometer.

Figure 1(a) shows the magnetization hysteresis loop measured at 5 K for the sample with an AlAs barrier thickness \da = 5~nm.  The magnetic field is applied along the easy axis of ``type-B" MnAs which is parallel to $[1\overline{1} 0]$ GaAs.\cite{tanakaAPL94,chunAPL01}  Two distinct transitions at 20 Oe and 500 Oe indicate the switching of magnetization direction of \GMA~ and MnAs, respectively; the coercive field of the \GMA~ layer is more readily seen by chemically removing the MnAs layer (see inset to Fig. 1(a)). Figure 1(b) shows the TMR for all 4 samples normalized at the high field value.  A sudden resistance drop is clearly seen when the magnetic moment of MnAs changes its direction from antiparallel to parallel with the \GMA~ magnetization.  We note that the TMR appears to show a non-monotonic dependence on the AlAs barrier thickness \da, with a striking effect of around 30$ \%$ for the sample with \da = 5 nm. The broad increase in background resistance coupled with the rotation of \GMA~ magnetization direction is probably due to the large difference in the magnitudes of magnetic moment.

The interpretation of these results is quite straightforward in analogy with metallic MTJs:  when the two ferromagnetic layers are magnetically aligned the tunneling probability is larger than when they are anti-aligned.  Quantitatively, the change in the tunnel resistance is given by \cite{julliere}:
\begin{equation}
TMR = \frac {\left( R_A - R_P \right)} {R_P} = \frac {2 P_G P_M}{\left( 1 - P_G P_M \right)}, 
\end{equation}
where \RA~ and \RP~ are the junction resistances with antiparallel and parallel moments, \PG~ and \PM~ are the spin polarizations of \GMA~ and MnAs respectively.  Hence, if the spin polarizations \PG~ and \PM~ are known, we can estimate the efficiency of spin polarized injection through the tunnel barrier as a ratio of the observed TMR to the ideal TMR predicted by the above equation. While direct measurements of \PG~ and \PM~ are not available at present,  band structure calculations predict 100 $\%$ spin-polarization in \GMA~ with $x \geq 0.125$ \cite{ogawaJMMM99} and for (hypothetical) zinc-blende  MnAs.\cite{shiraiJMMM98}  However, when MnAs is in the NiAs structure, it is not half-metallic and the theoretical value of the spin polarization is about 0.3.\cite{sanvitopc} If we assume that the half-metallicity of \GMA~ holds down to $x = 0.03$ (i.e. $P_G = 1$), the TMR observed in our measurements is close to 40 $\%$ of the ideal TMR.  We emphasize that this is a very conservative estimate since the bulk magnetization of \GMA \cite{potashnikPRB02} is far less than $4 \umu_B$ expected for a half-metallic system.\cite{ogawaJMMM99}  We can rule out spurious effects that might artificially enhance TMR in our measurements. First, the magnetoresistances of individual MnAs and \GMA~ layers are smaller than 0.5 $\%$ for the field range shown in Figure 1(b).  Further, the systematic variation of the TMR with barrier thickness rules out possible effects of fringe fields due to the nearby MnAs layer on \GMA. We now focus on the sample with \da =5 nm in order to examine the physics of these MTJs in some depth.

Unlike the case of all-metal MTJs, where there is a negligible change in magnetization with temperature, the magnetization and hence the spin-polarization of \GMA~ depend strongly on temperature.  Figure 2(a) shows the temperature dependence of the TMR  along with that of the bulk magnetization.  Both disappear at the Curie temperature of \GMA~ (around 70 K).  The figure also shows that the temperature dependences of the TMR and the magnetization are quite different. We propose two possible explanations for this difference: it may be related to the faster decay of surface magnetism, as is found in other half-metallic systems,\cite{parkPRL98} or it may indicate that the spin polarization in \GMA~ is not directly proportional to the magnetization. This correlation of the TMR with $T_C$ is a unique feature of these junctions where one can probe both ferromagnetic and paramagnetic states by changing the temperature.

The voltage dependence of the TMR at 4.2 K is shown in Fig. 2(b), along with the I-V characteristics at the same temperature.  We observe a very rapid decay of the TMR at voltages as low as 100 mV.  Although the behavior resembles that of metallic MTJs, the relative scale of the voltage is much smaller in these hybrid MTJs.\cite{mooderaPRL95} The decrease in the TMR with voltage may be attributed to the influence of the electric field  upon the tunnel barrier,\cite{bratkovskyPRB97} and implies low barrier heights for the semiconducting AlAs spacer compared to typical insulating Al$_2$O$_3$ spacers in metallic MTJs. Detailed information about the barrier can be obtained through the analysis of conductance-voltage (G-V) curves.  Figure 3 shows the G-V characteristics measured at zero magnetic field for several temperatures between 4.2 K and 240 K.  The large change of $G$ with voltage, especially at low temperatures, again indicates low barrier heights.  A distinct zero bias anomaly develops below the $T_C$ of \GMA, suggesting a small energy gap around the Fermi energy.\cite{hayashiJCG99} Another noticeable feature is the asymmetry in the G-V curves:  the conductance under positive bias (wherein the MnAs is at a higher potential) is larger than that under negative bias, and -- at temperatures above the Curie temperature of \GMA~ -- the minimum conductance occurs away from zero bias.  A full analysis of the G-V characteristics below the $T_C$ of \GMA~ is not possible because the detailed valence band structure of this material is presently not known from experiment.  Instead we focus on the G-V characteristics above $T_C$ where the conductance-voltage curves are parabolic within the voltage range $\pm 40$~mV (see for instance the data for $T \geq 120$K in Fig. 3).

The asymmetric shape and the occurrence of minimum conductance at a finite voltage lead us to apply the Brinkman-Rowell-Dynes (BDR) tunneling model that was originally developed to calculate the tunneling across metal-insulator-metal junctions with different barrier heights at the interfaces.\cite{BDR}  Although \GMA~ is not a metal, the BDR model is still applicable for voltages less than the Fermi energy of \GMA~ (0.16 eV if we assume a hole density of $1\times10^{20}$ cm$^{-3}$).  A best fit to the BDR model -- shown for the data at 120 K in Fig. 3 -- allows us to extract the barrier heights at the MnAs/AlAs interface ($\phi_1$) and the \GMA/AlAs interface ($\phi_2$), as well as the barrier thickness (\da) (see the inset of Fig. 3).

The best value for $\phi_1$~ is $0.15 \pm 0.01$ eV, which is much smaller than that obtained (0.8 eV) in studies of MnAs/AlAs/MnAs MTJs grown on (111) GaAs.\cite{tanakaAPL02}  This discrepancy can be  attributed to the different orientation of MnAs growth or to a subtle change at the interface.\cite{vanroyJCG01} We note that  measurements of Fe/GaN/Fe MTJs grown on (001) GaAs yield a small barrier height (0.11 eV), comparable to our results.\cite{nemethJCG01}  A simple estimate of $\phi_2$ is given  by the difference between the known AlAs-GaAs valence band offset (0.55 eV) and the Fermi energy of \GMA~ (0.16 eV).  Our result from fitting the data ($0.40 \pm 0.01$eV) is close to this estimate (0.39 eV).  Equally good agreement is found for the barrier thickness:  we  determine \da $= 6.7 \pm 0.1$~nm assuming light hole states participate in the tunneling through AlAs, while \da $= 4.4 \pm 0.1$~nm assuming heavy holes. It is quite possible that a mixture of light and heavy holes may provide a better description of reality, consistent with the designed AlAs thickness (5 nm). The successful application of the BDR model implies that the conduction is indeed due to tunneling processes.

The development of a zero bias anomaly below the Curie temperature of \GMA~ deserves further attention, since it may help in understanding the electronic structure of \GMA~ at the Fermi energy.  Figure 4(a) shows the conductance dip in the low bias conductance curves for parallel and anti-parallel spin orientations. As expected, the zero bias anomaly becomes more pronounced as the temperature is lowered from 4.2 K to 330 mK. Since \GMA~ is known to exhibit a metal-insulator transition, we analyze the behavior of the conductance-voltage characteristics within the context of early studies of disordered systems with a metal-insulator transition.\cite{altshulerSSC79,McMillanPRL81} On the metallic side of the metal-insulator transition, the one-electron density of states at the Fermi energy ($N(E)$) can be calculated using a scaling model that includes localization, correlation, and screening, and is given by:\cite{McMillanPRB81}
\begin{equation}
N(E) = N(0)[1+(E/\Delta)^{1/2}], 
\end{equation}
where the correlation gap $\Delta$ is a measure of the screening length.  As shown in Fig. 4(b), the conductance in the MTJs studied here is indeed proportional to $V^{1/2}$ except at the lowest bias where it is affected by thermal broadening.  Since the conductance is proportional to $N(E)$, a linear fit to the data in Fig. 4(b) yields $\Delta =$5.7 meV and $\Delta =$3.0 meV for parallel and anti-parallel configurations, respectively.  Surprisingly (and perhaps co-incidentally), these values of the correlation gap in \GMA~ are consistent with those extracted from studies of granular aluminum tunnel junctions\cite{DynesPRL81} in the regime wherein the Al conductivity is comparable to that of \GMA.  

In summary, we have shown that hybrid MnAs/AlAs/\GMA~ magnetic tunnel junctions provide an excellent model system for studying spin injection from a ferromagnetic metal into a semiconductor.  This is enabled by the observation of a large TMR whose magnitude tracks the magnetization of the \GMA~ layer. Modeling of the I-V characteristics at temperatures above the Curie temperature of the \GMA~ layer allow us to understand the nature of the barrier in great detail. Furthermore, analysis of the zero bias anomaly in conductance-voltage measurements shows a clear $V^{1/2}$ variation of the density of states, indicating strong electron correlation effects in \GMA~. 

N.S., S.H.C, and K.C.K were supported by grant nos. ONR N00014-99-1-0071, -0716, and DARPA/ONR N00014-99-1-1093.  P.S. and S.P. were supported by grant nos. DARPA N00014-00-1-0951 and NSF DMR 01-01318. We thank D. D. Awschalom and M. Flatte for a critical reading of this manuscript.

\newpage
\begin{center}
{\bf References}
\end{center}
%\bibliography{myrefs}

\newpage

\begin{center}
{\bf Figure Captions}
\end{center}

\begin{figure}[t]
\includegraphics{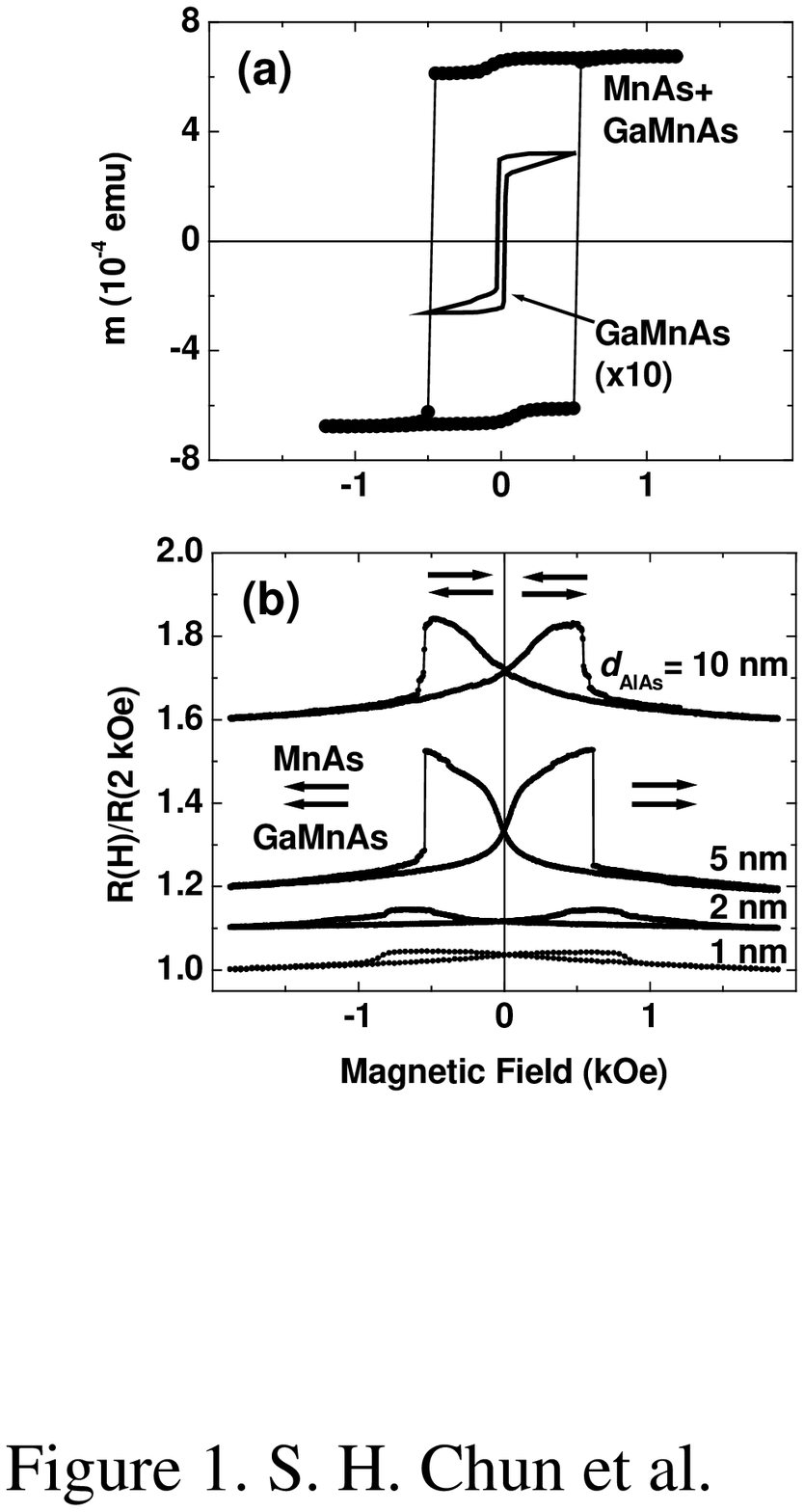}
\caption{Chun et al.\label{chunfig1}}
\end{figure}

FIG. \ref{chunfig1}:  (a) Magnetic hysteresis loops at $T=5$~Kfor a MnAs-\GMA~hybrid junction and for the same sample after the MnAs layer is removed. The transition of \GMA~ is broadened by the adjacent MnAs layer.  (b) Magnetoresistances of hybrid junctions with different AlAs barrier thickness ($T = 4.2$~K).  The curves are shifted for clarity.

\begin{figure}[t]
\includegraphics{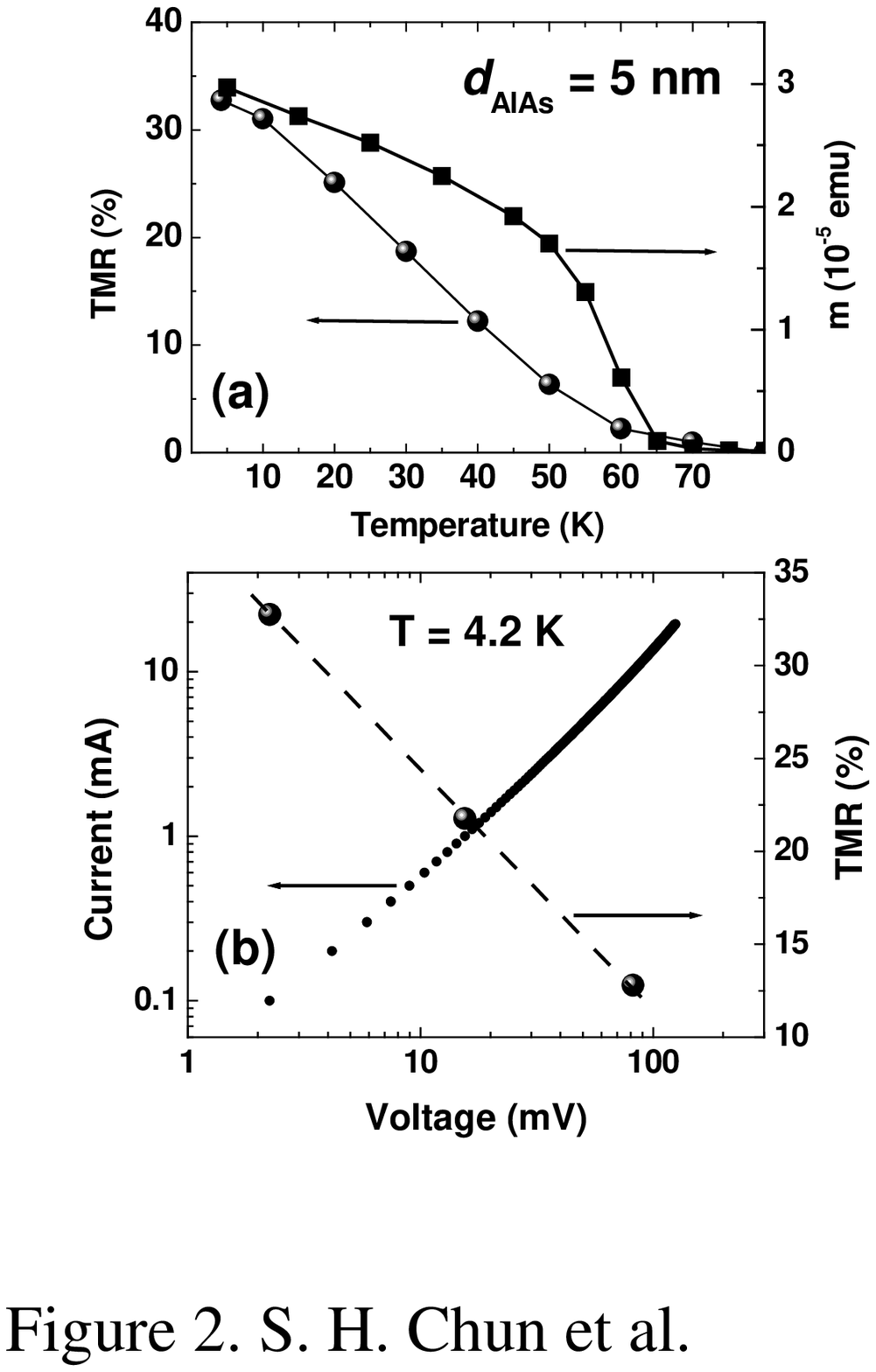}
\caption{Chun et al.\label{chunfig2}}
\end{figure}

FIG. \ref{chunfig2}:  (a) Temperature dependences of \GMA ~magnetization at 50 Oe and TMR measured with $I = 100 \umu$A for a junction with \da = 5 nm.  (b) I-V characteristics and voltage dependence of TMR for the same junction at 4.2 K.

\begin{figure}[t]
\includegraphics{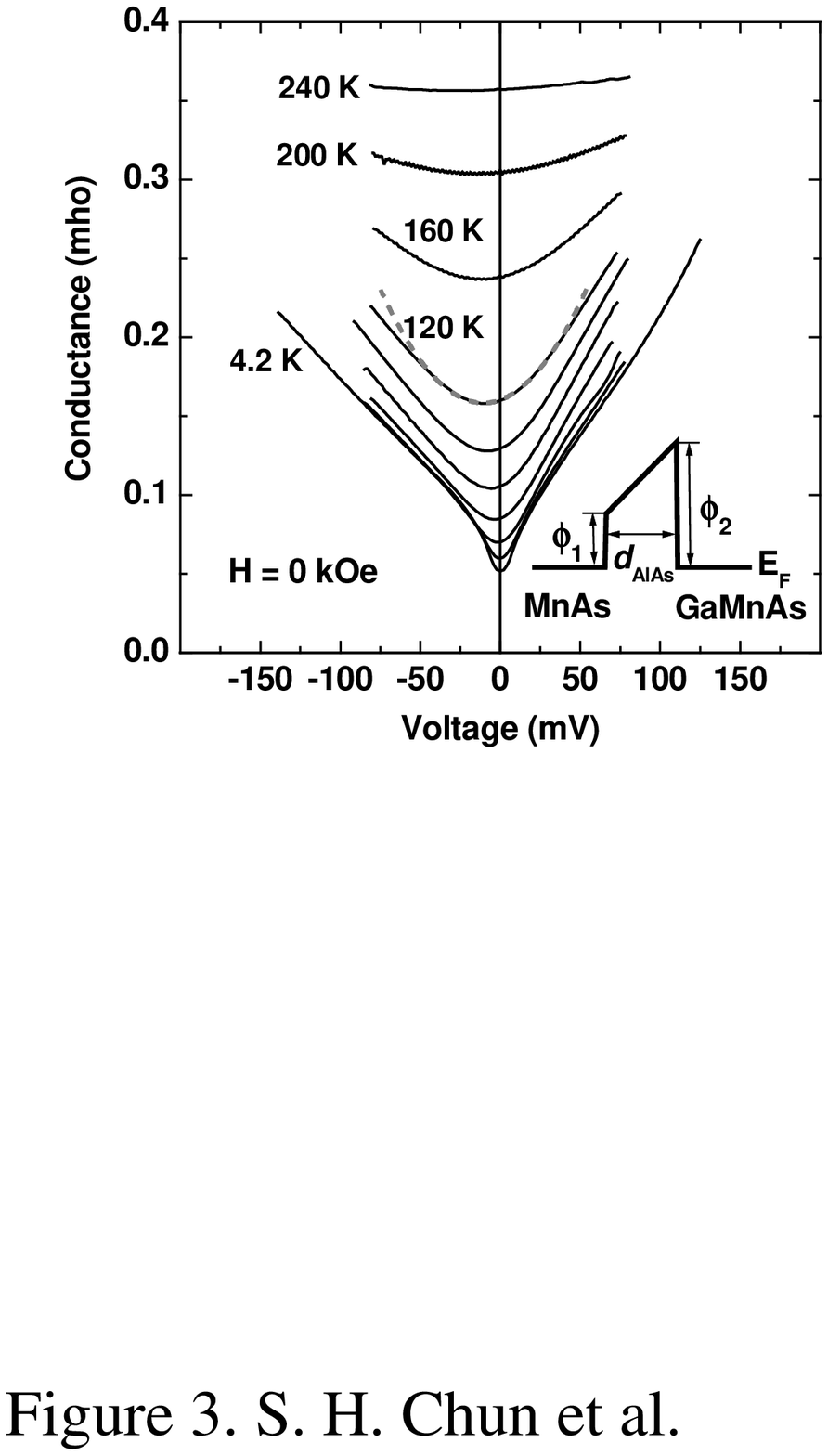}
\caption{Chun et. al.\label{chunfig3}}
\end{figure}

FIG. \ref{chunfig3}:  Zero field conductance curves of the junction used in Fig. 2 at selected temperatures.  $T = 20, 40, 60, 80,$ and 100 K for the curves between 4.2 K  and 120 K.  The dashed line superimposed on the 120 K data is a fit to Brinkman-Dynes-Rowell model over the range $\pm 40$~mV (see text).  The inset shows the schematic diagram of the model.

\begin{figure}[t]
\includegraphics{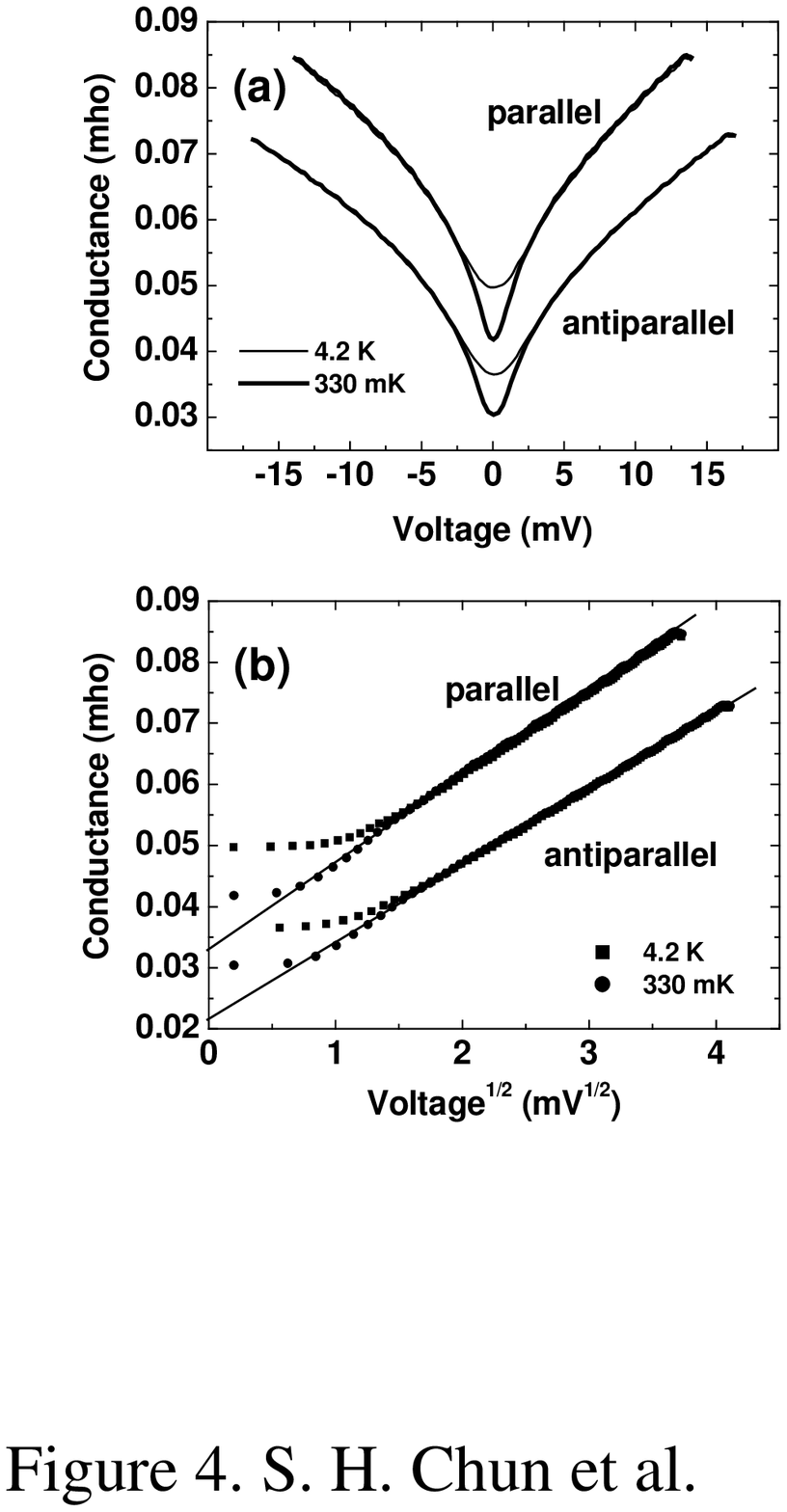}
\caption{Chun et. al.\label{chunfig4}}
\end{figure}

FIG. \ref{chunfig4}:  (a) Low bias conductance curves for the same junction as in Figs. 2 and 3 (\da = 5 nm) for parallel and antiparallel magnetization directions.  The zero-bias anomaly is more pronounced at 330 mK.  (b) The data in Fig. 4(a) plotted as a function of $V^{1/2}$.  Linear fits (solid lines) are used to extract the correlation gap (see text).  The deviation at low bias  is due to thermal broadening.
\clearpage
\pagestyle{empty}
\clearpage

\pagestyle{empty}

\clearpage

\clearpage

\end{document}